\documentclass[acs,preprint]{revtex4}

\usepackage{graphicx,amsmath,amsfonts,textcomp}

\newcommand{\fref}[1]{Fig.~\ref{#1}}

\begin{document}
\title{Conduction at domain walls in insulating Pb(Zr$_{0.2}$Ti$_{0.8}$)O$_3$ thin films}
\author{J. Guyonnet}
\altaffiliation{These authors contributed equally to the paper}
\affiliation{DPMC, University of Geneva, 24 Quai Ernest Ansermet,
1211 Geneva 4, Switzerland}
\author{I. Gaponenko}
\altaffiliation{These authors contributed equally to the paper}
\affiliation{DPMC, University of Geneva, 24 Quai Ernest Ansermet,
1211 Geneva 4, Switzerland}
\author{S. Gariglio}
\affiliation{DPMC, University of Geneva, 24 Quai Ernest Ansermet,
1211 Geneva 4, Switzerland}
\author{P. Paruch}
\affiliation{DPMC, University of Geneva, 24 Quai Ernest Ansermet,
1211 Geneva 4, Switzerland}
\date{\today}
\maketitle

 Ferroic domain walls occur naturally as intrinsically nanoscale interfaces separating different orientations of spontaneous electric polarisation, magnetization, or strain. They can present radically different properties from their parent materials, opening an alternative route towards unusual functionalities with enormous application potential \cite{bea_natmat_09_domainwalls_NV,salje_cpc_10_multiferroic_boundaries}. Much recent work has focused on the complex multiferroic insulator BiFeO$_3$ \cite{catalan_BFO_review}, with discoveries of possible magnetism in 109$^\circ$ domain walls \cite{martin_nl_08_xchange_BFO,daraktchiev_10_BFO_DW}, and conduction in 109$^\circ$, 180$^\circ$ \cite{seidel_natmat_09_BFO}, and subsequently 71$^\circ$ \cite{farokhipoor_condmat_DW_BFO} domain walls. To date, no clear consensus has emerged on the microscopic origin of the conduction \cite{seidel_natmat_09_BFO, lubk_prb_09_BFO_DW}, variously attributed to band gap lowering, or the presence of a potential step due to a polarisation discontinuity attracting charged defects. Both scenarios relate the conduction to the strongly non-Ising nature of BiFeO$_3$ domain walls \cite{kubel_AC_90_BFO}. Here, we demonstrate that 180$^\circ$ domain walls in the much simpler, purely ferroelectric tetragonal perovskite Pb(Zr$_{0.2}$Ti$_{0.8}$)O$_3$ (PZT) are also conducting. The observed domain wall conduction is clearly differentiable from polarisation switching currents and shows highly nonlinear and asymmetric voltage characteristics, strong temperature dependence, and high stability. 

At BiFeO$_3$ domain walls, oxygen vacancy defect states inside the band gap clearly modulate the conduction \cite{seidel_PRL_10_BFO_La}, but scanning tunneling spectroscopy measurements \cite{chiu_am_11_BFO_STS} also show effective band gap narrowing at 109$^\circ$ and 71$^\circ$ domain walls. A wide range of conduction mechanisms, both bulk- and interface-limited, have been suggested from local probe studies \cite{seidel_natmat_09_BFO,seidel_PRL_10_BFO_La,farokhipoor_condmat_DW_BFO}. To clarify the mechanism of domain wall conduction, measurements in a purely ferroelectric, tetragonal perovskite system such as PZT, without the complexities and multiferroic nature of BiFeO$_3$, are extremely useful. Interestingly, PZT and PbTiO$_3$ show little change in band gap values between ferroelectric (3.6 eV) and paraelectric (3.4 eV) phases \cite{bilc_prb_08_ferroelectric_DFT}, suggesting that band gap narrowing would not play a significant role. Even the ``simple'' 180$^\circ$ domain walls in this material can show a complex internal structure, with polarisation rotation and flux-closure under specific electric or strain boundary conditions \cite{aguado_PRL_08_closure,lee_prb_09_nonising_DW}. 

Recent transmission electron microscopy measurements beautifully demonstrate such polarisation rotation in a narrow region at the junction of a 180$^\circ$ domain wall and the interface of a PZT film and its 1.5-unit-cell-thick SrRuO$_3$ substrate \cite{jia_sci_11_quadrant_DWs}. Such features present a potential step necessitating screening, which could occur via charged defects, providing donor/trap states near the ferroelectric surface and facilitating electron emission/tunneling from an adjacent electrode. Alternatively, internal charge redistribution in semiconducting films could locally increase conductivity at the wall. Mean-field models considering 180$^\circ$ domain walls in uniaxial ferroelectric semiconductors show that charging due to domain wall inclination significantly increases static conductivity \cite{eliseev_condmat_DW_inclined}. Although energetically less favourable, such inclined domain walls can be present during polarisation switching, and can be locally stabilized within thin films by the presence of charged defects \cite{jia_natmat_08_charged_DW}. We note that the dynamic nature of the domain wall conduction in BiFeO$_3$ was highlighted by recent measurements suggesting that microscopically irreversible distortions of the polarisation structure are at its origin, and could be present in other ferroelectric domain walls \cite{maksymovych_nl_11_BFO_DW_conductivity}. Local probe studies in PZT have demonstrated a strong electroresistance effect, with conduction across thin films controlled by polarisation orientation \cite{maksymovych_sci_09_TER_PZT} in agreement with theoretical predictions \cite{tsymbal_sci_06}. Essentially, the ferroelectric thin film with its adjoining electrodes (one of which can be the mobile metallic atomic force microscope (AFM) tip) is seen as two back-to-back Schottky diodes, with conduction dominated by interface or bulk limited contributions depending on specific material parameters (dielectric response, mobility of charge carriers, presence of donor/trap states, metal and ferroelectric work functions and electrode geometry) \cite{zubko_jap_06_PZT,pintilie_prb_07_PZTconduction,maksymovych_nanotech_11_PZT_conduction}. However, no observations of domain-wall-specific conduction have previously been reported in this material.

All our measurements were carried out in ultra high vacuum on epitaxial PZT thin films grown on SrRuO$_3$ bottom electrodes on SrTiO$_3$ substrates. As described previously \cite{gariglio_apl_07_PZT_highTc, guyonnet_apl_09_shear}, the films show high crystalline and surface quality. The polarisation axis is perpendicular to the film plane, and all films are monodomain (up-polarised) as-grown. In these films we created 180$^\circ$ domain walls by switching the ferroelectric polarisation using a biased scanning AFM tip. The resulting domains were imaged by piezoresponse force microscopy (PFM) to determine the polarisation orientation, and by conductive-tip AFM (c-AFM) for local current maps.

\begin{figure}
\includegraphics[width=\columnwidth]{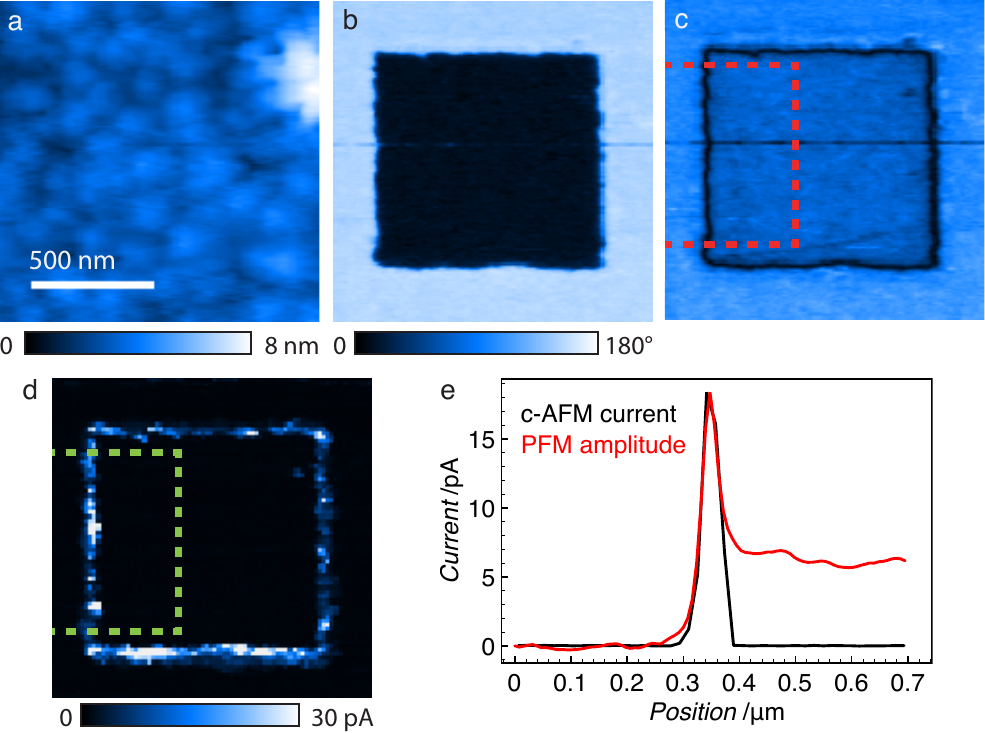}
\caption{Domain wall conduction in PZT. Topography (a) showing an rms roughness of 0.4 nm. PFM phase (b) and amplitude (c) images of a square domain written with a positive tip bias. (d) c-AFM measurement at -1.5 V tip bias, showing an average current value of 25 pA at the position of domain walls. (e) Line profiles of c-AFM current and inverse PFM amplitude averaged over the left domain wall, as indicated by the dashed lines in (c) and (d).}
\label{square}
\end{figure}

As shown in \fref{square} for a 70 nm thick film, under a low negative DC tip bias we observe current signal at the domain walls, while the bulk of the sample remains insulating within our experimental resolution. In fact, as can be seen from the average signal profiles in \fref{square}(e), the effective domain wall width measured by PFM (convoluted with the nominally $\sim$45 nm radius tip) exactly corresponds to the lateral extent of this current signal. No correlation between the regions of high current in the domain walls and the morphology of the film was detected, and no features corresponding to the domain wall position were present in the surface topography. 

\begin{figure*}
\includegraphics[width=0.95\textwidth]{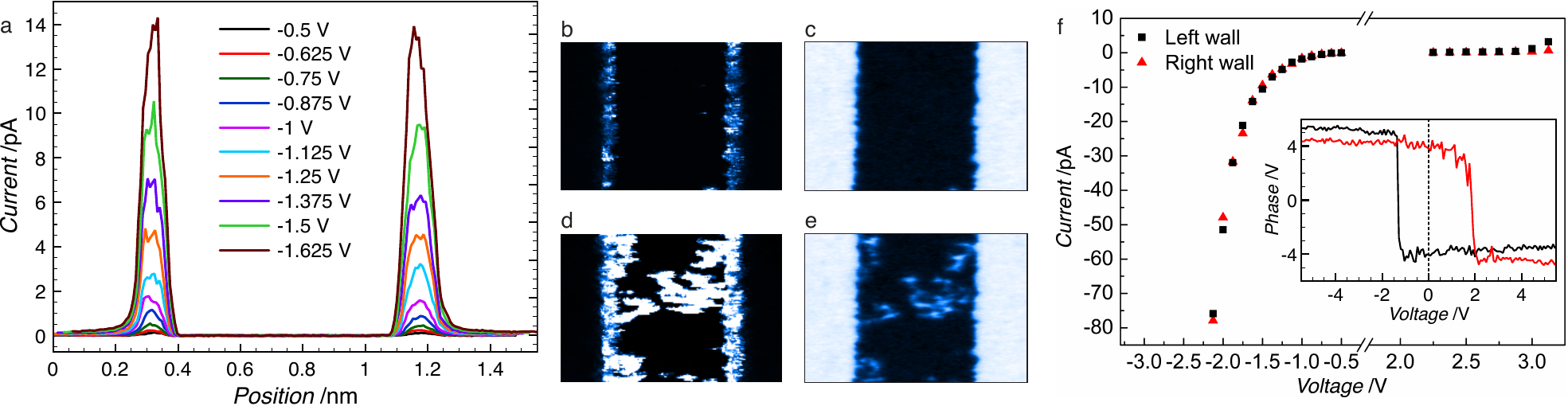}
\caption{Voltage dependence of domain wall conduction. (a) Current values from successive c-AFM scans with increasing tip bias, for sub-switching voltages from -0.50 V to -1.625 V, averaged over 120 scan lines. (b) -1.625 V c-AFM and (c) subsequent PFM measurement. (d) -2.25 V c-AFM measurement showing switching centers with current values on the order of 2--4 nA and (e) subsequent PFM measurement, demonstrating corresponding partial switching. (f) Average domain wall current as a function of the applied voltage with PFM measurements of polarisation hysteresis under applied voltage in inset.}
\label{I_V}
\end{figure*}

To better understand the microscopic mechanisms underlying the observed domain wall current ($I$), we investigated its dependence on the applied voltage ($V$). On stripe domains written in the same film, we carried out alternating PFM and c-AFM measurements with a stepwise increase of 0.125V for each iteration of the latter. As shown in \fref{I_V}(a), for negative tip voltages, we first detect current at -0.5 V. From the lack of any evolution of the ferroelectric domain structure imaged by PFM, it is clear that the observed current is not due simply to polarisation reversal and domain wall motion in the applied electric field (\fref{I_V}(b,c)). Moreover, static measurements at low voltages applied for 120 s likewise show persistent domain wall current over 120 s with no change in the domain configuration (see Supplementary Materials). In fact, although displacement current associated with polarisation switching is unambiguously seen at higher voltages (beyond -2.2 V), it is of much greater magnitude, and accompanied by changes in the domain configuration (\fref{I_V}(d,e)). Combining these data with similar measurements for positive tip bias, we can see that the $I$--$V$ characteristics of the domain wall are strongly non-linear and asymmetric (\fref{I_V}(f)). For positive bias, we do not observe domain wall current until much higher voltages (2.25 V). We observe a similarly asymmetric polarisation switching hysteresis (inset) and relate these to the non-identical electrodes leading to different effective Schottky barrier heights. 

\begin{figure}
\includegraphics[width=0.85\columnwidth]{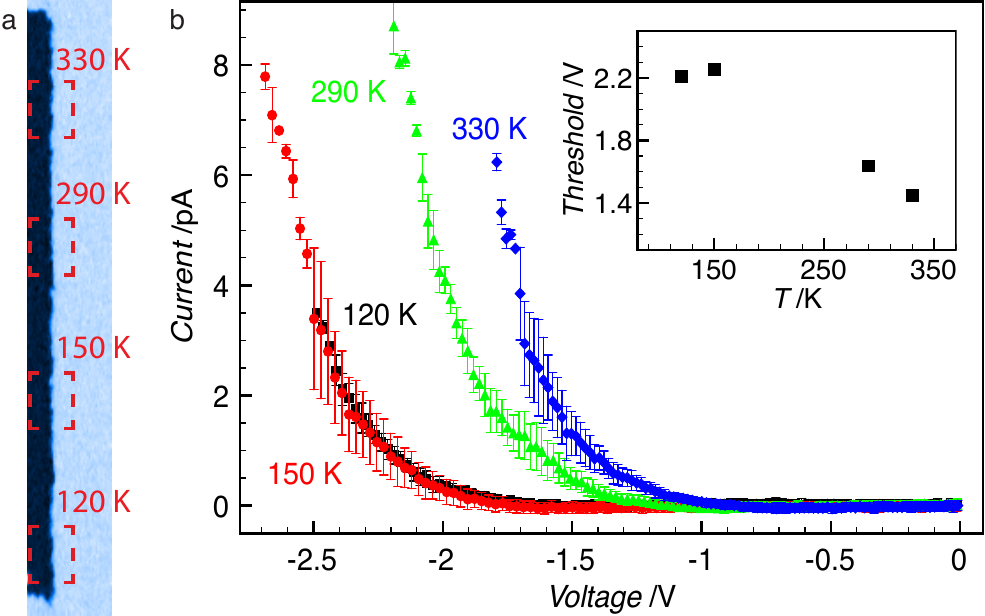}
\caption{Domain wall conduction at different temperatures. (a) PFM image of domain structure with $I$-$V$ measurement regions indicated by dashed lines. (b) Average current as a function of applied voltage, with the inset showing conduction thresholds ($I \geq 1.0$ pA) for the different temperatures.}
\label{temperature}
\end{figure}

A more complete data set for the $I$--$V$ characteristics was obtained from dynamic grid measurements on designated segments of the domain wall at different temperatures, as indicated in \fref{temperature}(a). For each 16 $\times$ 16 nm$^2$ sector of the grid, the voltage was swept from zero to $V_{max}$ and back to zero on the stationary tip, with increasing $V_{max}$ for each subsequent measurement on the same grid. As shown in \fref{temperature}(b), the $I$--$V$ characteristics averaged along the grid sectors containing the domain wall appear almost identical at 120 and 150 K. At higher temperature, however, thermal activation is evident, with increasing current and decreasing conduction threshold voltage (\fref{temperature}(b) inset). More details of the measurements can be found in Supplementary Materials. 

We analyzed these data within the framework of different mechanisms including interface-limited conduction via Schottky thermionic emission (SE) $\log(I/T^2) \propto e/(k_B T)(\sqrt{eE/(4\pi\varepsilon_0\varepsilon)} - \phi_B)$ \cite{Sze_semiconductors} or Fowler-Nordheim tunneling (FN) $\log (I/E^2) \propto -1/E$ \cite{fowler_prsa_28_FN_tunneling} and bulk-limited conduction via Poole-Frenkel hopping (PF) $\log (I/E) \propto (-U_{trap}/(k_B T))(\sqrt{eE/(4\pi\varepsilon_0\varepsilon)})$ or space charge limited conduction (SCL) $I \propto V^n$. In these expressions $e$ is the electronic charge, $T$ the temperature, $E$ the electric field, $k_B$ the Boltzmann constant, $\varepsilon$ the dielectric constant, $\varepsilon_0$ the vacuum permittivity, $\phi_B$ the barrier height, and $U_{\rm trap}$ the trap energy. For SCL, different values of the exponent $n$ are expected ($n=2$ for discrete traps, $n>2$ with $(n-1) \propto 1/T$ for traps distributed within the band gap) \endnote{For more detailed consideration of these mechanisms applied to conduction across ferroelectric PZT thin films please refer to \cite{zubko_jap_06_PZT,pintilie_prb_07_PZTconduction,maksymovych_nanotech_11_PZT_conduction}.}. The electric field $E$ across an insulating ferroelectric film is uniform for a parallel plate capacitor, but inhomogeneous in the case of an AFM tip, although a linear dependence on voltage can still be demonstrated ($E \propto V$). In a semiconducting ferroelectrics parallel plate capacitor the presence of free charge, in particular screening polarisation near the film surface, gives a maximum field $E = \sqrt{\frac{2qN_D}{\varepsilon_{DC}\varepsilon_0}(V + V_{bi} - \frac{k_bT}{q})}$ in this region, with electronic transport limited primarily by the reverse-bias Schottky barrier at the film-electrode interface (abrupt junction approximation) \cite{Sze_semiconductors}, where $N_D$ is the dopant density, $q$ the carrier charge, $\varepsilon_{DC}$ the static dielectric constant and $V_{bi}$ the resulting in-built bias. In the following discussion, we consider $E\propto V$. Analysis of the data in the abrupt junction approximation yields unphysically high dopant concentration values (see Supplementary Materials).

\begin{figure}
\includegraphics[width=\columnwidth]{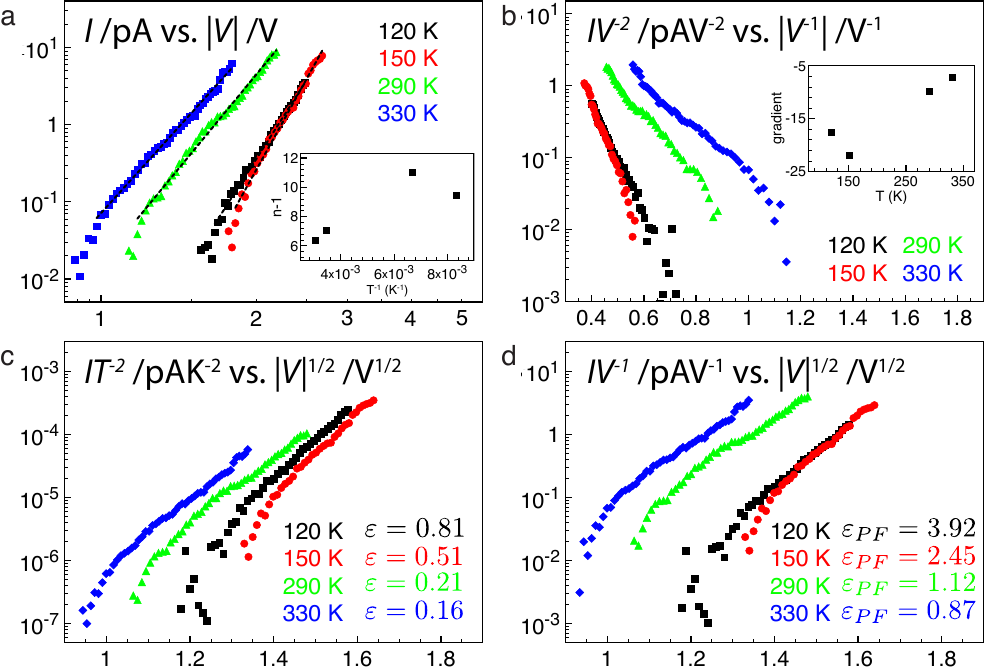}
\caption{Analysis of $I$--$V$ characteristics for different conduction mechanisms. (a) linearisation in $I \propto V^n$ (SCL), with $(n-1)$ vs. $1/T$ in inset. (b) linearisation in $I/V^2$ vs. $1/V$ (FN), with the gradient vs. $T$ in inset. (c) linearisation $I/T^2$ vs. $\sqrt{V}$ (SE). (d) linearisation in $I/V$ vs. $\sqrt{V}$ (PF).}
\label{analysis}
\end{figure}

The high $n$ values of the $I \propto V^n$ fits of \fref{analysis}(a), and lack of obvious linearisation in $n-1$ vs. $1/T$ suggest that SCL, either in the discrete trap limit or with a continuous trap distribution, is unlikely.  However, representations of the data in normalised coordinates corresponding to FN, SE and PF all show reasonable linearisation (\fref{analysis}(b--d)). Very intense fields near the film surface, as modeled numerically in Supplementary Materials, would promote electron tunneling or emission into this region, suggesting an interface-limited mechanism in agreement with the significantly increased conduction when negative bias is applied to the tip rather than the bottom SrRuO$_3$ electrode. 
However, the obvious variation of the $\log (I/V^2)$ vs. $1/V$ gradient with temperature (inset) suggests that FN, for which this quantity should be temperature-independent, does not adequately describe the data, while fits of $\log(I)$ vs. $\sqrt{V}$ in the SE regime yield unphysically low dielectric constant values, expected to be in the $[\varepsilon_\infty = 6;\varepsilon_{DC} = 400]$ range \cite{zubko_jap_06_PZT}.  More realistic $\varepsilon$ values are obtained from the $\log (I/V)$ vs. $\sqrt{V}$ fits in the PF regime, although these are still slightly smaller than the optical dielectric constant $\varepsilon_\infty$. Although our data preclude a quantitative Arrhenius analysis of the $\log(I)$ vs. $1/T$ dependence expected in this scenario, the almost identical $I$-$V$ response observed at 120 and 150 K suggests more complex features. One possibility would be tunneling-assisted Poole-Frenkel conduction, previously observed in HfO$_2$ thin films \cite{jeong_jap_05_HFOconduction}, where tunneling/emission from the highly asymmetric electrodes gives rise to the diode-like $I$-$V$ characteristics, but conduction within the domain wall itself proceeds by Poole-Frenkel hopping.

\begin{figure}
\includegraphics[width=0.9\columnwidth]{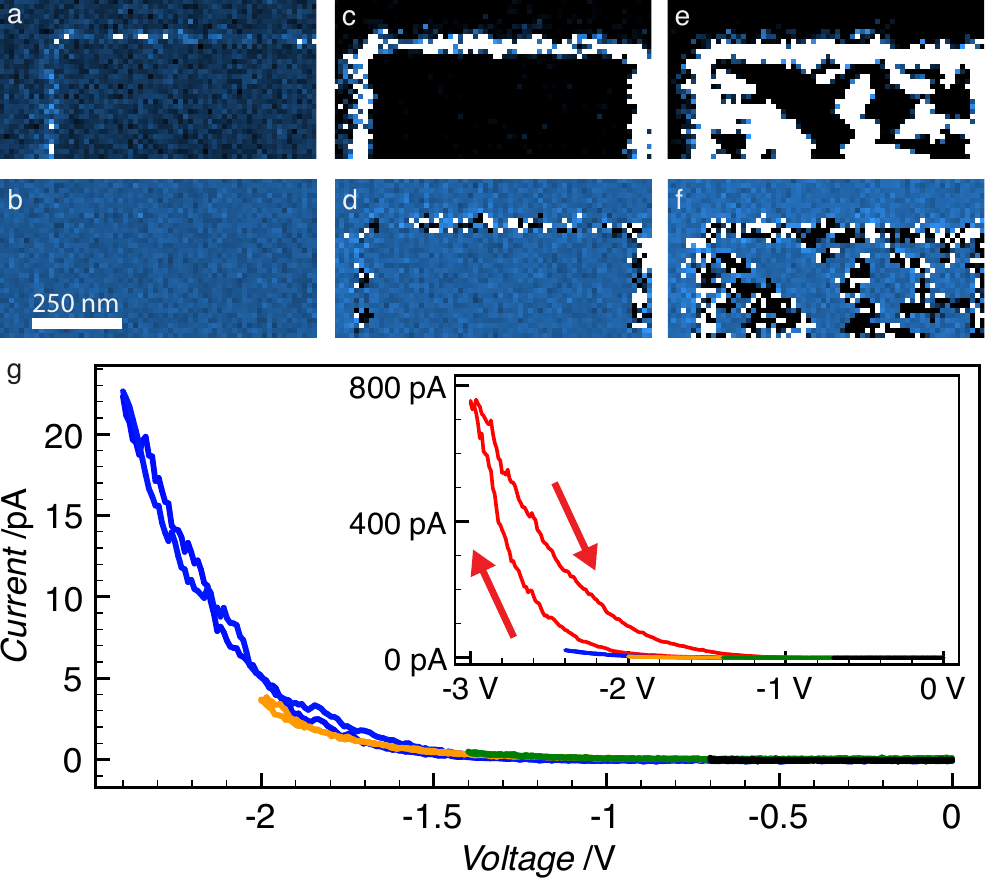}
\caption{Current-voltage hysteresis at domain walls. (a,c,e) C-AFM images of the domain wall conduction at $V_{\rm max} =$ -1.4, -2.0, and -2.4 V. (b,d,f) Local hysteresis for voltage sweeps up to $V_{\rm max} =$ -1.4, -2.0, and -2.4 V. (g) Average $I$-$V$ domain wall response for the domain wall in the sub-switching regime: for the domain wall as a whole, no hysteresis is observed until voltages close to the switching threshold are reached. Inset shows the strong $I$-$V$ hysteresis once polarisation at switching voltages.}
\label{hysteresis}
\end{figure}

To further explore the nature of the domain wall conduction, we also compared the current response for forward and reverse directions of voltage sweeps in the grid measurements carried out to progressively higher voltage $V_{\rm max}$, observing three distinct regimes. Initially, domain wall conduction with no hysteresis (calculated locally for each $16\times16$ nm$^2$ pixel as $\int I_{\rm forward}-I_{\rm reverse}$) is detected (at -0.9 to -1.4 V in these measurements). Subsequently (-1.5 to -2.7 V), local hysteresis variations are observed (\fref{hysteresis}(c,d,g)) with no changes in domain structure. These can be positive (higher current during forward voltage sweep) or negative (higher current during reverse voltage sweep) and vary randomly for consecutive voltage sweeps, including within the same grid sector, giving little to no hysteresis in the average $I$-$V$ response, although negative hysteresis begins to dominate at the upper part of the range. Finally, upon reaching polarisation switching (beyond -2.8 V) we observe much higher current values, and strong negative hysteresis both at the domain walls and within the domains (\fref{hysteresis}(e,f,g, inset)), as expected for irreversible changes in the polarisation structure. 

Taken together, these data point to a thermally activated, domain-wall-specific phenomenon. As for conduction through thin ferroelectric films, asymmetric Schottky barriers and electrode tunneling into trap states near the interface appear to play a key role. However, there is clearly not enough trap density to support conduction across the ferroelectric film outside the domain walls. The latter must therefore act as a local trap reservoir, for which a range of activation energies could be envisaged. Such behaviour can be well correlated with the microscopic domain wall structure detected by TEM, with polarisation rotation near the film-electrode interface and small, apparently randomly spaced 1--2 unit cell steps throughout the length of the wall presenting maximally anti-aligned dipole moments \cite{jia_sci_11_quadrant_DWs}. These charged domain wall segments necessitate screening, and would thus strongly increase defect density at the wall. Defects such as oxygen vacancies could provide the necessary trap states for electronic tunneling from the electrode into the ferroelectric layer and conduction via subsequent hopping between trap states, and would be expected to show strong thermal activation. Rather than the uniform static conductivity predicted for inclined domain walls \cite{eliseev_condmat_DW_inclined}, we would expect significant local variations as a result of the microscopic structure - as are indeed observed - depending on both the density of charges/defects and the local connectivity of the charge regions around the domain wall steps. 

The lack of hysteresis at low bias suggests a strong pinning of the domain wall and accompanying trap distribution in their initial metastable state, with predominantly conductive domain wall currents. At higher voltages, the microscopic structure of the domain wall could locally evolve at lengthscales smaller than the domain wall thickness/PFM resolution limit, with local hysteresis, different from the observations of global negative hysteresis through most of the conductive regime in BiFeO$_3$ \cite{maksymovych_nl_11_BFO_DW_conductivity}. At high enough fields the depinning regime with large scale domain wall motion and reaccommodation of defects/screening charges accompanying polarisation switching would result in a new metastable configuration. A crucial and surprising result in this scenario is the very high stability of the domain wall conduction over time. Although inclined 180$^\circ$ domain walls would thermodynamically be expected to relax to an elastically ideal straight configuration, leading to a decrease in the conductive signal, we observe persistent domain wall current over 120 s at low voltages, and the measured current levels at a given domain wall segment remain stable when sampled over up to 4 days at 295 K. We also note that in such PZT films, our previous measurements have always demonstrated very high stability of even 10-20 nm domains over weeks or months and at elevated temperatures \cite{paruch_apl_06_stability}, suggesting a high persistence of the configuration, presumably due to very effective defect-controlled pinning of the domain walls \cite{paruch_prl_05_dw_roughness_FE,paruch_jap_06_dynamics_FE}. This result is particularly encouraging for potential device applications, for which local conductive paths created using such domain walls should not deteriorate over time.

In conclusion, our results clearly demonstrate conduction at 180$^\circ$ domain walls in ferroelectric tetragonal PZT, strongly suggesting that this phenomenon is far more general than has previously been shown. The scenario of defect segregation at partially charged domain walls could be present in a number of ferroelectric materials, and potentially controlled by strain and electrostatic boundary conditions, opening interesting possibilities for future research. In particular, studies looking for domain wall conduction in a large range of ferroelectric materials should be envisaged to identify the best candidates for potential devices. The possibility of tuning the observed response by doping or oxygen vacancy engineering can also be explored. From a fundamental point of view, it would be interesting to compare the behaviour of domain walls in thin films and in single crystals, and to combine PFM, c-AFM and TEM studies, in particular to investigate the effects of intrinsic domain walls structure on the conduction.

\subsection{Methods}
60-70 nm PZT films on 30-40 nm metallic SrRuO$_3$ were epitaxially grown on single-crystal (001) SrTiO$_3$ substrates by off-axis radio-frequency magnetron sputtering. X-ray diffraction analysis showed relaxiation, with a $c$-axis lattice parameter values of $\sim$4.16 \AA. AFM topography measurements showed 4 {\AA} rms surface roughness.

PFM and c-AFM measurements were carried out in an {\it Omicron VT} beam deflection AFM under ultra-high vacuum ($\sim3\times 10^{-10}$ mbar) using a {\it Nanonis RC4} real-time controller and {\it Bruker} CoCr-coated MESP tips on samples introduced from ambient with no subsequent treatment. Typical scan size and rates were 3 $\mu$m and 600 nm/s. For PFM, a 2 V, 1 kHz AC tip bias was applied, and PFM signal measured with an {\it SRS SR830} lock-in amplifier (10 ms time constant). For c-AFM, a 0--10 V DC tip bias was applied, and current measured using an {\it Omicron PRE E} preamplifier with a $\pm500$ fA noise level. Conduction threshold values depend on tip condition, with $\sim$400 mV for the best tips.

\subsection*{}
The authors thank P. Zubko and J.-M. Triscone for helpful discussions, and M. Lopes and S. Muller for technical support. This work was funded by the Swiss National Science Foundation through the NCCR MaNEP, and by the European Commission FP7 project OxIDes.

The authors declare that they have no competing financial interests.

Correspondence should be addressed to P.P.~(email: patrycja.paruch@unige.ch).

All authors discussed the results and interpretation presented. S. G. grew and characterized the PZT thin films. J. G. and I. G. carried out all PFM and c-AFM measurements and wrote the paper with P. P. 

\end{document}